\documentclass[aps,prev,twocolumn,preprintnumbers,floatfix,nofootinbib]{revtex4-2}
\usepackage{graphicx}
\usepackage[ngerman,english]{babel}
\usepackage{bm}
\usepackage{times}
\usepackage{slashed}
\usepackage{mathtools}
\usepackage{color}
\usepackage{epsfig}
\usepackage{dcolumn}
\usepackage{textcomp}
\usepackage{color}
\usepackage{mathtools}
\usepackage{hyperref}
\hypersetup{pdfstartview=FitV,colorlinks=true,linkcolor=blue,citecolor=blue,filecolor=black,urlcolor=blue}

\begin{document} 


\title{Rare Kaon Decay to Missing Energy: Implications of the NA62 Result for a $Z^\prime$ Model}

\author{T\'essio B. de Melo$^1$}\email{tessiomelo@gmail.com}
\author{Sergey Kovalenko$^2$}\email{sergey.kovalenko@unab.cl}
\author{Farinaldo S. Queiroz$^{1,3}$}\email{farinaldo.queiroz@iip.ufrn.br}
\author{C. Siqueira$^{1,4}$}\email{csiqueira@ifsc.usp.br}
\author{Yoxara S. Villamizar$^{1,3}$}\email{yoxara@ufrn.edu.br}

\affiliation{$^1$ International Institute of Physics, Universidade Federal do Rio Grande do Norte, Campus Universitario, Lagoa Nova, Natal-RN 59078-970, Brazil\\
$^2$ Departamento de Ciencias F\'isicas, Universidad Andres Bello, Sazie 2212, Santiago, Chile\\
$^3$ Departamento de F\'isica, Universidade Federal do Rio Grande do Norte, 59078-970, Natal,
RN, Brasil\\
$^4$ Instituto de F\'isica de S\~ao Carlos, Universidade de S\~ao Paulo, Av. Trabalhador S\~ao-carlense 400, S\~ao Carlos, Brasil.}

\begin{abstract}
\noindent
Meson decays offer a good opportunity to probe new physics. The rare kaon decay $K^+ \rightarrow \pi^+ \nu\bar{\nu}$ is one of the cleanest of them and, for this reason, is rather sensitive to new physics, in particular, vector mediators. NA62 collaboration, running a fixed-target experiment at CERN, recently reported an unprecedented sensitivity to this decay, namely a branching fraction of $BR(K^+ \rightarrow \pi^+ \nu\bar{\nu}) = (11^{+4.0}_{-3.5})\times 10^{-11}$ at 68\% C.L. Vector mediators that couple to neutrinos may yield a sizeable contribution to this decay. Motivated by the new measurement, we interpret this result in the context of a concrete $Z^\prime$ model, and put our findings into perspective with the correlated $K_L \rightarrow \pi^0 \nu \bar{\nu}$ decay measured by KOTO collaboration, current, and future colliders, namely the High-Luminosity and High-Energy LHC.
\end{abstract}

\maketitle
\flushbottom

\section{Introduction}
\label{sec_introduction}
\begin{figure}[h]
    \centering
    \includegraphics[scale=0.27]{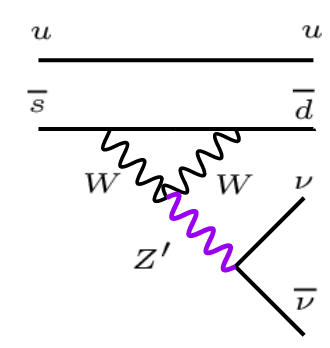}
    \includegraphics[scale=0.2]{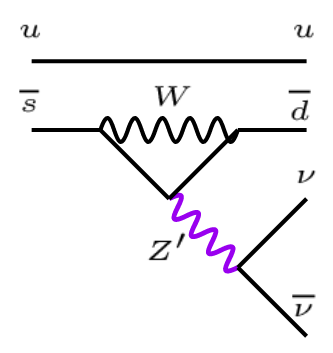}
    \caption{Feynman diagrams that illustrate how vector mediators can contribute to the $K^+ \rightarrow \pi^+ \nu\bar{\nu}$ decay. The first diagram requires $Z^\prime$ coupling to neutrinos, whereas the second further requires couplings to the top quark.}
    \label{ClarissaGigante} 
\end{figure}

Mesons have played a key role in the understanding of properties of elementary particles.  The introduction of strangeness along with isospin lead us to the eight-fold way, based on the SU(3) flavor symmetry. These theoretical insights have contributed to the discovery of quantum chromodynamics as we know today. Another good example is the famous $\theta-\tau$ puzzle. Two different decays were found for charged strange mesons known at the time as $\theta^+$ and $\tau^+$. The decay modes had different parities, but the particles were shown to have the same mass and lifetime. It was indeed a puzzling observation. Later, it was realized that weak interactions violate parity, and these two particles were actually the same $K^+$-meson. Additionally, the  Glashow-Iliopoulos-Maiani (GIM) mechanism and quark charm surfaced as an explanation of the absence of weak flavor changing neutral currents in the processes such as  $K^+ \rightarrow \pi^+ \nu\bar{\nu}$.  The discovery of CP violation in the $K^0-\bar{K}^0$ system further proved the importance of meson physics for our understanding of nature. Furthermore, meson systems are able to access possible new sources of CP violation that are of paramount importance for explaining the observed baryon-antibaryon asymmetry in our universe \cite{Dine:2003ax}. Lastly, the $K^+$ rare decay into neutrinos can efficiently probe the presence of heavy vector mediators, beyond the Standard Model (SM), \cite{Cogollo:2012ek,Queiroz:2016gif,Kitahara:2019lws,Borah:2020swo,Dutta:2020scq,Jho:2020jsa,Aebischer:2020mkv,Kang:2020gfi} via the Feynman diagrams displayed in Fig.\ref{ClarissaGigante}. 

The meson decay $K^+ \rightarrow \pi^+ \nu\bar{\nu}$ is a flavor changing neutral current process that occurs in the SM via the box and penguin diagrams, with the latter being dominated by the top quark contribution.
Due to the GIM and loop suppression, the SM contribution to this decay is very small, reading  ${\rm BR}( K^+ \rightarrow \pi^+ \nu\bar{\nu}) =  (8.4 \pm 1.0) \times 10^{-11}$ \cite{Buras:2015qea}, while the NA62 currently imposes ${\rm BR}( K^+ \rightarrow \pi^+ \nu\bar{\nu}) = 11.0^{+4.0}_{-3.5} \times 10^{-11}$  \cite{NA62last,CortinaGil:2020vlo}  (results collected in 2016, 2017 and 2018). Therefore, one can clearly notice that the current sensitivity achieved by NA62 is not far from the SM prediction. Having said that, the NA62 collaboration has been continuously searching for the rare kaon decay \cite{CortinaGil:2020vlo}. KOTO collaboration has also  conducted some searches as well, but offering weaker constraints \cite{Ahn:2020opg}\footnote{KOTO collaboration has also recently reported the
observation of three anomalous events in the $K_L \rightarrow \pi^0 \nu\bar{\nu}$. This anomaly requires the branching ratio for this decay mode to be about two orders of magnitude larger than the SM one \cite{Kitahara:2019lws}, indicating the presence of a new light and long-lived particle with mass of the order of $100$~MeV. There is no such light particle in our model. Hence the KOTO anomaly will be regarded as an statistical fluke.}. To concretely show the relevance of the recent NA62 result, we will put it into perspective other existing limits in a model that features a heavy vector mediator. The model is based on the $SU(3)_C \times SU(3)_L \times U(1)_Y$ gauge group, known as 3-3-1 model. It is well motivated by the ability to naturally explain the observed replication of the generations, and nicely hosting dark matter candidates \cite{Mizukoshi:2010ky,Kelso:2013nwa,Profumo:2013sca,Cogollo:2014jia,Montero:2017yvy,Carvajal:2017gjj,Abdalla:2019xka,Leite:2020bnb}, addressing neutrino masses \cite{Vien:2018otl,Nguyen:2018rlb,CarcamoHernandez:2019iwh,CarcamoHernandez:2019vih}, among other interesting phenomena \cite{Dong:2015dxw,Dong:2015dxw,Alves:2016fqe,Queiroz:2016gif,Dong:2017zxo,Huong:2019vej,Arcadi:2019uif,VanLoi:2020xcq,Duy:2020hhk,Dias:2020ryz}. As a result of the enlarged gauge group, the 3-3-1 model have several new gauge bosons, such as $W^\prime$ and $Z^\prime$ bosons, which are subject to restrictive constraints rising from collider physics \cite{Cao:2016uur,Nepomuceno:2019eaz,Nepomuceno:2019goa,Liu:2011at}, muon anomalous magnetic moment \cite{Kelso:2013zfa,Kelso:2014qka,deJesus:2020ngn,deJesus:2020upp}, and lepton flavor violation \cite{Arcadi:2017xbo,Ferreira:2019qpf}. We will investigate the role of the $Z^\prime$ gauge boson in the rate $K^+$ decay and then use this to draw bounds on the $Z^\prime$ mass using the $K_L \rightarrow \pi^0 \bar{\nu}\nu$, $K^+ \rightarrow \pi^+ \bar{\nu}\nu$ decays. 

Our work is structured as follows: In {\it Section} \ref{sec_model}, we review the 3-3-1 model and compute the $Z^\prime$ couplings necessary to our analyses; in {\it Section} \ref{sec:kaon}, we discuss the computation of the branching fractions of the kaons in our model; in {\it Section} \ref{sec:res}, we discuss our results and, finally in {\it Section} \ref{sec:conc}, we draw our conclusions. 

\section{The Model}
\label{sec_model}
The 3-3-1 models are extensions of the standard model and are based on the following local symmetry group: $\textbf{SU(3)}_\textbf{C} \times \textbf{SU(3)}_\textbf{L} \times \textbf{U(1)}_\textbf{N}$, where \textbf{C} corresponds to the color charge, \textbf{L} denotes the left-handed fermions and \textbf{N} is the quantum number of the $\text{U(1)}$ group. The general expression for the electric charge operator in these models is written as,
\begin{equation}\frac{Q}{e}=\frac{1}{2}\left(\lambda_{3}+\beta \lambda_{8}\right)+\text{N} \,  \text{I}=\left(\begin{array}{c}
 \frac{1}{3}+\text{N}\\
-\frac{2}{3}+\text{N} \\
 \frac{1}{3}+\text{N}
\end{array}\right),\label{charged}\end{equation}
where $\lambda_{3}=\operatorname{diag}(1,-1,0)$, $\lambda_{8}=\operatorname{diag}(1,1,-2) / \sqrt{3}$, and $\text{I}$ are the diagonal Gell-Mann matrices and the identity matrix, respectively. We took $\beta=-\frac{1}{\sqrt{3}}$ because in our work we choose the model known as 3-3-1 with right-handed neutrinos (3-3-1 r.h.n)~\cite{Hoang:1996gi,Hoang:1995vq}. However, we highlight that our conclusions are also applicable to the 3-3-1 model with neutral fermions proposed in \cite{Mizukoshi:2010ky}. The hypercharge in this model is given as,
\begin{equation}Y=2Q-\lambda_3 = 2 N-\frac{\sqrt{3} \lambda_{8}}{3},\label{hych}\end{equation}
which is identical to the standard model one. 
The left (L) and right (R)-handed fermionic fields of this model are represented as follows,

\begin{equation}
\label{eq:Lepton-Triplet-1}\begin{array}{c}
f_{L}^{a}=\left(\begin{array}{c}
\nu_{L}^{a} \\
\ell_{L}^{a} \\
\left(\nu_{R}^{c}\right)^{a}
\end{array}\right) \sim(1,3,-1 / 3),\\ \\ \ell_{R}^{a} \sim(1,1,-1),\end{array}\end{equation}

\begin{equation}
\label{eq:Quark-Triplet-1} 
\begin{array}{c}
Q_{i L}=\left(\begin{array}{c}
d_{i L} \\
-u_{i L} \\
d_{i L}^{\prime}
\end{array}\right) \sim(3, \overline{3}, 0),\\ \\
u_{i R} \sim(3,1,2 / 3),\, d_{i R} \sim(3,1,-1 / 3),\\ \\ d_{i R}^{\prime} \sim(3,1,-1 / 3), \\ \\
Q_{3 L}=\left(\begin{array}{c}
u_{3 L} \\
d_{3 L} \\
T_{L}
\end{array}\right) \sim(3,3,1 / 3), \\ \\
u_{3 R} \sim(3,1,2 / 3),\, d_{3 R} \sim(3,1,-1 / 3),\\ \\ T_{R} \sim(3,1,2 / 3),
\end{array}\end{equation}
where $a=1,2,3$ and $i=1,2$ are the generation indexes, $f^a_L$ and $Q_{i L}$, $Q_{3L}$  represent the left-handed leptonic and quark triplets, respectively. These fields encompass the SM spectrum like neutrinos ($\nu^a=\nu_e, \nu_\mu, \nu_\tau$), charged leptons $\ell^a=e, \mu, \tau $, and quarks $u_i=\overline{u},\overline{c}$, $d_i=\overline{d},\overline{s}$, $u_3=t$ and $d_3=b$. Besides, there are other particles additional  to the SM: the right-handed neutrino $\left(\nu_{R}^{c}\right)^{a}$ and three new heavy exotic quarks $d_{i L}^{\prime}$ and $T_{L}$. In Eqs.
(\ref{eq:Lepton-Triplet-1}), (\ref{eq:Quark-Triplet-1}), we have specified the field assignments indicating how they transform under the symmetries $\left(\mathrm{SU}(3)_{c}, \mathrm{SU}(3)_{L}, \mathrm{U}(1)_{N}\right)$, respectively.
The values of their electric charge and hypercharge can be found from Eqs. \eqref{charged} and \eqref{hych}.

Furthermore, the 3-3-1 r.h.n model contains three scalar fields $\chi$, $\eta$ and $\rho$ in the following representations
\begin{eqnarray}
\quad \quad  \chi=\left(\begin{array}{c}
\chi^{0} \\
\chi^{-} \\
\chi^{\prime 0}
\end{array}\right) \sim(1,3,-1 / 3), \nonumber\\  \nonumber\\ 
\rho=\left(\begin{array}{c}
\rho^{+} \\
\rho^{0} \\
\rho^{\prime +}
\end{array}\right) \sim(1,3,2 / 3), \\ \nonumber\\ 
\eta=\left(\begin{array}{c}
\eta^{0} \\
\eta^{-} \\
\eta^{\prime 0}
\end{array}\right) \sim(1,3,-1 / 3).\nonumber \label{scalars}
\end{eqnarray}

These scalar triplets in  Eq. \eqref{scalars} are responsible for the spontaneous symmetry breaking (SSB) in the model, with the following vacuum (vev) structure,  
\begin{equation} \langle\chi\rangle=\left(\begin{array}{c}
0 \\
0 \\
v_{\chi}
\end{array}\right), \langle\rho\rangle=\left(\begin{array}{c}
0 \\
v_{\rho} \\
0
\end{array}\right),\langle\eta\rangle=\left(\begin{array}{c}
v_{\eta} \\
0 \\
0
\end{array}\right),\end{equation}.

We will assume that $v_{\chi}\gg v_{\eta},v_{\rho}$, leading to the two-step SSB, 
\begin{align*}
    \textbf{SU(3)}_\textbf{L} \times \textbf{U(1)}_\textbf{X}  \xrightarrow{\quad \langle\chi\rangle \quad} \textbf{SU(2)}_\textbf{L} \times \textbf{U(1)}_\textbf{Y} \xrightarrow{ \langle\eta\rangle, \langle\rho\rangle } \textbf{U(1)}_\textbf{Q}
\end{align*}with $U(1)_Q$, being the $U(1)$ from electrodynamics.

The fermion masses rise from the Yukawa Lagrangian that reads,
 \begin{equation}\begin{aligned}
\mathcal{L}_{Y u k} &=\lambda_{1 a} \bar{Q}_{1 L} d_{a R} \rho+\lambda_{2 i a} \bar{Q}_{i L} u_{a R} \rho^{*}+G_{a b}^{\prime} \bar{f}_{L}^{a} e_{R}^{b} \rho\\ &+G_{a b} \varepsilon^{i j k}\left(\bar{f}_{L}^{a}\right)_{i}\left(f_{L}^{b}\right)_{j}^{c}\left(\rho^{*}\right)_{k}+\lambda_{3 a} \bar{Q}_{1 L} u_{a R} \eta\\ &+\lambda_{4 i a} \bar{Q}_{i L} d_{a R} \eta^{*}+\lambda_{1} \bar{Q}_{3 L} T_{R} \chi+\lambda_{2 i j} \bar{Q}_{i L} d_{j R}^{\prime} \chi^{*} + H . c .
\end{aligned}\end{equation}

The quark and charged lepton masses are proportional to $v=246$~GeV, where $v^{2}=v_{\rho}^{2}+v_{\eta}^{2}$ similarly to the SM. The fourth term leads to a $3 \times 3$ antisymmetric neutrino mass matrix \cite{Hoang:1995vq}, which means that the model has one massless and two degenerate neutrino mass eigenstates. Moreover, the gauge bosons $W$ and $Z$ acquire mass term identical to the SM as well. In addition to the SM fields, there are new massive gauge bosons predicted by the model as result of the enlarged gauge symmetry, denoted as $Z^{\prime}, V^{\pm}$ and $U^{0}, U^{0 \dagger}$. The masses of these fields are,
\begin{equation}\begin{aligned}
M_{W^{\pm}}^{2} &=\frac{1}{4} g^{2} v^{2}, M_{Z}^{2}=\frac{M_{W^{\pm}}^{2}}{C_{W}^{2}} \\
M_{Z^{\prime}}^{2} &=\frac{g^{2}}{4\left(3-4 S_{W}^{2}\right)}\left[4 C_{W}^{2} v_{\chi}^{2}+\frac{v^{2}}{C_{W}^{2}}+\frac{v^{2}\left(1-2 S_{W}^{2}\right)^{2}}{C_{W}^{2}}\right] \\
M_{V^{\pm}}^{2} &=\frac{1}{4} g^{2}\left(v_{\chi}^{2}+v^{2}\right), M_{U^{0}}^{2}=\frac{1}{4} g^{2}\left(v_{\chi}^{2}+v^{2}\right).
\end{aligned}\end{equation}
with $M_{W} \ll M_{U}, M_{V}$, $S_W=\sin \theta_W$ and $C_W=\cos \theta_W$, with $\theta_W$, the Weinberg angle. 

The charged (CC) and neutral (NC) currents are found to be,

 \begin{equation}\begin{aligned}
		\mathcal{L}^{C C}=&-\frac{g}{\sqrt{2}}\left[\bar{\nu}_{L}^{a} \gamma^{\mu} e_{L}^{a} W_{\mu}^{+}+\left(\nu_{R}^{\bar{c}}\right)^{a} \gamma^{\mu} e_{L}^{a} V_{\mu}^{+}\right.\\
		&+\left.\bar{\nu}_{L}^{a} \gamma^{\mu}\left(\nu_{R}^{c}\right)^{a} U_{\mu}^{0}\right]\\ &-\frac{g}{\sqrt{2}}\left[\left(\bar{u}_{3 L} \gamma^{\mu} d_{3 L}+\bar{u}_{i L} \gamma^{\mu} d_{i L}\right) W_{\mu}^{+}\right. \\ &+\left.\left(\bar{T}_{L} \gamma^{\mu} d_{3 L}+\bar{u}_{i L} \gamma^{\mu} d_{i L}^{\prime}\right) V_{\mu}^{+}\right.\\
		&+\left(\bar{u}_{3 L} \gamma^{\mu} T_{L}-\bar{d}_{i L}^{\prime} \gamma^{\mu} d_{i L}\right) U_{\mu}^{0}+\text { h.c. }],
		\label{lcc}
\end{aligned}\end{equation}

\begin{equation}\begin{aligned}
		\mathcal{L}^{N C} &=\frac{g}{2 c_{W}}\left\{\bar{f} \gamma^{\mu}\left[a_{1 L}(f)\left(1-\gamma_{5}\right)+a_{1 R}(f)\left(1+\gamma_{5}\right)\right] f Z_{\mu}^{1}\right.\\
        \label{eq:NC-Lagrangian}
		&\left.+\bar{f} \gamma^{\mu}\left[a_{2 L}(f)\left(1-\gamma_{5}\right)+a_{2 R}(f)\left(1+\gamma_{5}\right)\right] f Z_{\mu}^{2}\right\},
\end{aligned}\end{equation}
The second and third term in Eq. \eqref{lcc} violate leptonic number and weak isospin \cite{Hoang:1995vq}. $Z^1$ and $Z^2$ are neutral physical gauge bosons, which rise from the $Z$ and $Z^\prime$ gauge boson mixtures. $a_{1 R}(f)$, $a_{1 L}(f)$, $a_{2 R}(f)$ and $a_{2 L}(f)$ are couplings of fermions with the $Z^1$ and $Z^2$ bosons. The mixing angle of these bosons is commonly denoted as $\phi$ and when $\phi=0$, the couplings of $Z^1$ with the leptons and quarks are the same as the boson $Z$ in the SM. Likewise,
the couplings of $Z^2$ in this limiting case should be the same as $Z^\prime$ \cite{Hoang:1995vq}.
%
 These couplings for the vertices $Z
^\prime-\nu-\overline{\nu}$ , $Z^\prime-\overline{d_{i}}-d_{i}$ and $Z^\prime-\overline{b}-b$  are shown in Table \ref{tab1}.

\begin{table}[!htp]
\vspace{5mm}
    \centering
    \begin{tabular}{|c||c||c||c|} \hline
    & $Z
^\prime-\nu-\overline{\nu}$& $Z^\prime-\overline{d_{i}}-d_{i}$ & $Z^\prime-\overline{b}-b$ \\ \hline \hline
& & & \\
Coupling constant& $\frac{1-2 S_W^2}{2\sqrt{3-4S_W^2}}$& $-\frac{\sqrt{3-4S_W^2}}{6}$ & $\frac{3-2S_W^2}{6\sqrt{3-4S_W^2}}$\\
& & & \\ \hline
    \end{tabular}
    \caption{$Z
^\prime$ couplings to neutrinos and the left-handed down-type quarks, considering  $\phi=0$.}
    \label{tab1}
\end{table}

In order to study meson physics in our model, and investigate its connection to the $Z^\prime$ boson, we need to extract flavor changing neutral current. To do so, we start by reviewing how the flavor changing terms arise. The quark fields the following standard rotation are,
\[
\left(\begin{array}{l}
u \\
c \\
t
\end{array}\right)_{L}=V_{L}^{u}\left(\begin{array}{c}
u^{\prime} \\
c^{\prime} \\
t^{\prime}
\end{array}\right)_{L},\left(\begin{array}{l}
d \\
s \\
b
\end{array}\right)_{L}=V_{L}^{d}\left(\begin{array}{l}
d^{\prime} \\
s^{\prime} \\
b^{\prime}
\end{array}\right)_{L},
\] where $V_{L}^{u}$ and $V_{L}^{d}$ are the $3 \times 3 $ unitary matrices such that for the Cabibbo-Kobayashi-Maskawa (CKM) matrix we have $V_{\mathrm{CKM}} \equiv V_{L}^{u\dagger} V_{L}^{d}$.
Note that only the left-chiral terms of the Lagrangian (\ref{eq:NC-Lagrangian}) lead to the quark flavor violation. The right-chiral quark couplings to $Z'$  in Eq. (\ref{eq:NC-Lagrangian}) are independent of flavor and therefore are flavor-diagonal in the mass eigenstate basis. We can write these terms in the form,

\begin{equation}
    \mathcal{L}_{Z^\prime} \supset \frac{g}{C_W} (\overline{D_L^\prime} \gamma^\mu Y^D_L D_L^\prime) Z^\prime_\mu,
\end{equation}with $D_L^\prime = (d^\prime,s^\prime,b^\prime)^T$, and 
\begin{eqnarray}
Y^D_L = &&\frac{1}{6\sqrt{3-4S_W^2}}\times \\
\nonumber
&&\mathrm{Diagonal}(-3+4S_W^2,-3+4S_W^2,3-2S_W^2).
\end{eqnarray}
Changing the basis we get,
\begin{equation}
    \mathcal{L}_{Z^\prime} \supset \frac{g}{C_W} (\overline{D_L} \gamma^\mu Y^{D\prime}_L D_L) Z^\prime_\mu =  \Delta^{sd}_L(\overline{s_L} \gamma^\mu d_L) Z^\prime_\mu + ...,
\end{equation}where $D_L^\prime = V_L^d D_L$ and $Y^{D\prime}_L = V_L^{d\dagger} Y^D_L V_L^{d}$. Using the unitarity of the $V_L$ matrix, we finally find the coupling between the quarks $d$ and $s$, which is given by, 
\begin{equation}
\label{zp_s_d_coup}
    \Delta_L^{sd} = \frac{gC_W}{3-4S_W^2}V_{L32}^* V_{L31},
\end{equation}

Analogously for the neutrino-$Z'$ coupling we have,
\begin{equation}
\label{zp_nu_coup}
    \Delta_L^{\nu\bar{\nu}} = \frac{g}{2C_W}\frac{1-2 S_W^2}{\sqrt{3-4S_W^2}}.
\end{equation}

In principle, we can vary the entries of the matrix $V_L^d$ freely since the CKM matrix does not constrain $V^d_L$, but the product $V^d_L V^u_L$.
%
So, we choose the following parametrization for the $V_L^d$ matrix \cite{Buras:2012dp}
\begin{widetext}
\begin{equation}
V_L^d=\left(\begin{array}{ccc}
\tilde c_{12}\tilde c_{13} & \tilde s_{12}\tilde c_{23} e^{i \delta_3}-\tilde c_{12}\tilde s_{13}\tilde s_{23}e^{i(\delta_1
-\delta_2)} & \tilde c_{12}\tilde c_{23}\tilde s_{13} e^{i \delta_1}+\tilde s_{12}\tilde s_{23}e^{i(\delta_2+\delta_3)} \\
-\tilde c_{13}\tilde s_{12}e^{-i\delta_3} & \tilde c_{12}\tilde c_{23} +\tilde s_{12}\tilde
\tilde s_{13}\tilde s_{23}e^{i(\delta_1-\delta_2-\delta_3)} & -\tilde s_{12}\tilde s_{13}\tilde c_{23}e^{i(\delta_1 -\delta_3)}
-\tilde c_{12}\tilde s_{23} e^{i \delta_2} \\
-\tilde s_{13}e^{-i\delta_1} & -\tilde c_{13}\tilde s_{23}e^{-i\delta_2} & \tilde c_{13}\tilde c_{23}
\end{array}\right).
\label{VL-param}
\end{equation}
\end{widetext}
where $\tilde{s}_{ij} = \sin{\tilde{\theta}_{ij}}$, $\tilde{c}_{ij} = \cos{\tilde{\theta}_{ij}}$ and $\delta_i$ are the phases, with $i,j = 1,2,3$. For our purposes, the following entries will be important,
\begin{eqnarray}
    V_{L31}^d &=& -\tilde s_{13}e^{-i\delta_1} \\
    V_{L32}^d &=& -\tilde c_{13}\tilde s_{23}e^{-i\delta_2}
\end{eqnarray}
then, the product which appears in the $\Delta_L^{sd}$ coupling is
\begin{equation}
    V_{L31}^d V_{L32}^{d*} = -\tilde s_{13}\tilde c_{13}\tilde s_{23} e^{-i(\delta_1 - \delta_2)} \equiv |V_{L32}^{d*}V_{L31}^d|e^{-i\delta}
\end{equation}
where we leave the product $|V_{L32}^{d*}V_{L31}^d|$ and the phase $\delta$ as free parameters.

\section{Kaon decays}
\label{sec:kaon}

The rare Kaon decay modes $ K ^+ \to \pi ^+ \nu \bar{\nu} $ and $ K _L \rightarrow \pi ^0 \nu \bar{\nu} $ are considered golden modes in flavor physics, as they are very well understood theoretically and are sensitive to new physics contributions. In the SM both decays occur only at loop level and are dominated by $Z$ penguin and box diagrams. The corresponding branching ratios have been calculated 
at a high precision, including NNLO QCD, electroweak corrections and also non-perturbative and isospin breaking effects \cite{Buchalla:1997kz, Brod:2010hi, Buras:2005gr, Buras:2006gb, Brod:2008ss}.

The decay $ K ^+ \rightarrow \pi ^+ \nu \bar{\nu} $ is CP conserving, whereas $ K _L \rightarrow \pi ^0 \nu \bar{\nu} $ is CP violating. In the 331 model, the new sources of flavor and CP violation which contribute to these decays come from the $Z ^\prime$ interactions with ordinary quarks and leptons, as discussed above. Although these couplings induce the transitions at tree level, they are suppressed by the large $Z ^\prime$ mass.



Following the notation of Ref. \cite{Buras:2013ooa}, we can write the branching ratios for the Kaon decay modes $ K \to \pi \nu \bar{\nu} $ as,
\begin{equation}
\begin{split}
& BR ( K ^+ \rightarrow \pi ^+ \nu \bar{\nu} ) = \\
& \text{\ \ \ \ \ \ \ \ \ \ \ \ \ } \kappa _+ \left[ \left( \frac{\text{Im} X _{eff}}{\lambda ^5} \right) ^2 + \left( \frac{\text{Re} X _{eff}}{\lambda ^5} - \bar{P} _c (X) \right) ^2 \right] ,
\label{eq:br_kp}
\end{split}
\end{equation}
and,
\begin{equation}
BR ( K _L \rightarrow \pi ^0 \nu \bar{\nu} ) = \kappa _L \left( \frac{\text{Im} X _{eff}}{\lambda ^5} \right) ^2 .
\label{eq:br_kL}
\end{equation}
In these expressions $\lambda$ denotes the Cabibbo angle, $ \kappa _+ $ and $ \kappa _L $ are given by,
\begin{equation*}
\kappa _+ = (5.21 \pm 0.025) \cdot 10 ^{-11} \left( \frac{\lambda}{0.2252} \right) ^8 ,
\end{equation*}
\begin{equation*}
\kappa _L = (2.25 \pm 0.01) \cdot 10 ^{-10} \left( \frac{\lambda}{0.2252} \right) ^8 ,
\end{equation*}
and $P _c (X)$ summarizes the charm contribution,
\begin{equation*}
\bar{P} _c (X) = \left( 1 - \frac{\lambda ^2}{2} \right) P _c (X) ,
\end{equation*}
with,
\begin{equation*}
P _c (X) = (0.42 \pm 0.03) \left( \frac{0.2252}{\lambda} \right) ^4 .
\end{equation*}
$ X _{\text{eff}} $ describes the contribution of short distance physics,
\begin{equation*}
X _{\text{eff}} = V _{ts} ^* V _{td} X _L (K) ,
\end{equation*}
where,
\begin{equation}
\label{sm_331_contr}
X _L (K) = \eta _X X _0 (x _t) + \frac{\Delta _L ^{\nu \bar{\nu}}}{g _{SM} ^2 m _{Z ^\prime} ^2} \frac{\Delta _L ^{sd}}{V _{ts} ^* V _{td}} ,
\end{equation}
and,
\begin{equation*}
g _{SM} ^2 = 4 \frac{m _W ^2 G _F ^2}{2 \pi ^2} = 1.78137 \times 10 ^{-7} \text{\ GeV} ^{-2} .
\end{equation*} 

The first term in Eq. (\ref{sm_331_contr}) 
represents the SM contribution, which is dominated by $Z$-penguin and box diagrams, and includes QCD and electroweak corrections. The factor $\eta _X$ is close to unity, $ \eta _X = 0.994 $, and
\begin{equation*}
X _0 (x _t) = \frac{x _t}{8} \left[ \frac{x _t + 2}{x _t - 1} + \frac{3 x _t - 6}{(x _t - 1) ^2} \ln x _t \right] ,
\end{equation*}
with $ x _t = m _t ^2 / m _W ^2 $.
The 331 contribution is enclosed in the second term of Eq. (\ref{sm_331_contr}), with $ \Delta _L ^{sd} $ and $ \Delta _L ^{\nu \bar{\nu}} $ given in the Eqs. (\ref{zp_s_d_coup}) and (\ref{zp_nu_coup}), respectively. If $Z ^\prime$ is absent we have $ \Delta _L ^{\nu \bar{\nu}} = \Delta _L ^{sd} = 0 $ and the SM result is recovered. 
%
%
\\

The decays $K ^+ \rightarrow \pi ^+ \nu \bar{\nu}$ and $K _L \rightarrow \pi ^0 \nu \bar{\nu}$ are related to each other in the SM, via isospin symmetry, since both transitions are ruled by the same short distance operator. This interdependence leads to the Grossman-Nir limit \cite{Grossman:1997sk},
\begin{equation}
BR ( K _L \rightarrow \pi ^0 \nu \bar{\nu} ) \leq 4.3 BR ( K ^+ \rightarrow \pi ^+ \nu \bar{\nu} ) .
\end{equation}
For a fixed value of $BR ( K ^+ \rightarrow \pi ^+ \nu \bar{\nu} )$, this theoretical bound provides an upper limit for $BR ( K _L \rightarrow \pi ^0 \nu \bar{\nu} )$, which is typically still stronger than the current experimental limits. The bound remains valid in SM extensions in which the new physics is much heavier than the Kaon mass. In particular, the 331 model obeys the bound, as we will show in the next section. \\

Having presented the formulas that summarizes the predictions of the 331 model regarding meson FCNC processes, we can now discuss the implications of the experimental searches for the flavor violating Kaon decays on the parameter space of the model.

\begin{figure*}[t]
    \centering
    \includegraphics[width=0.48\textwidth]{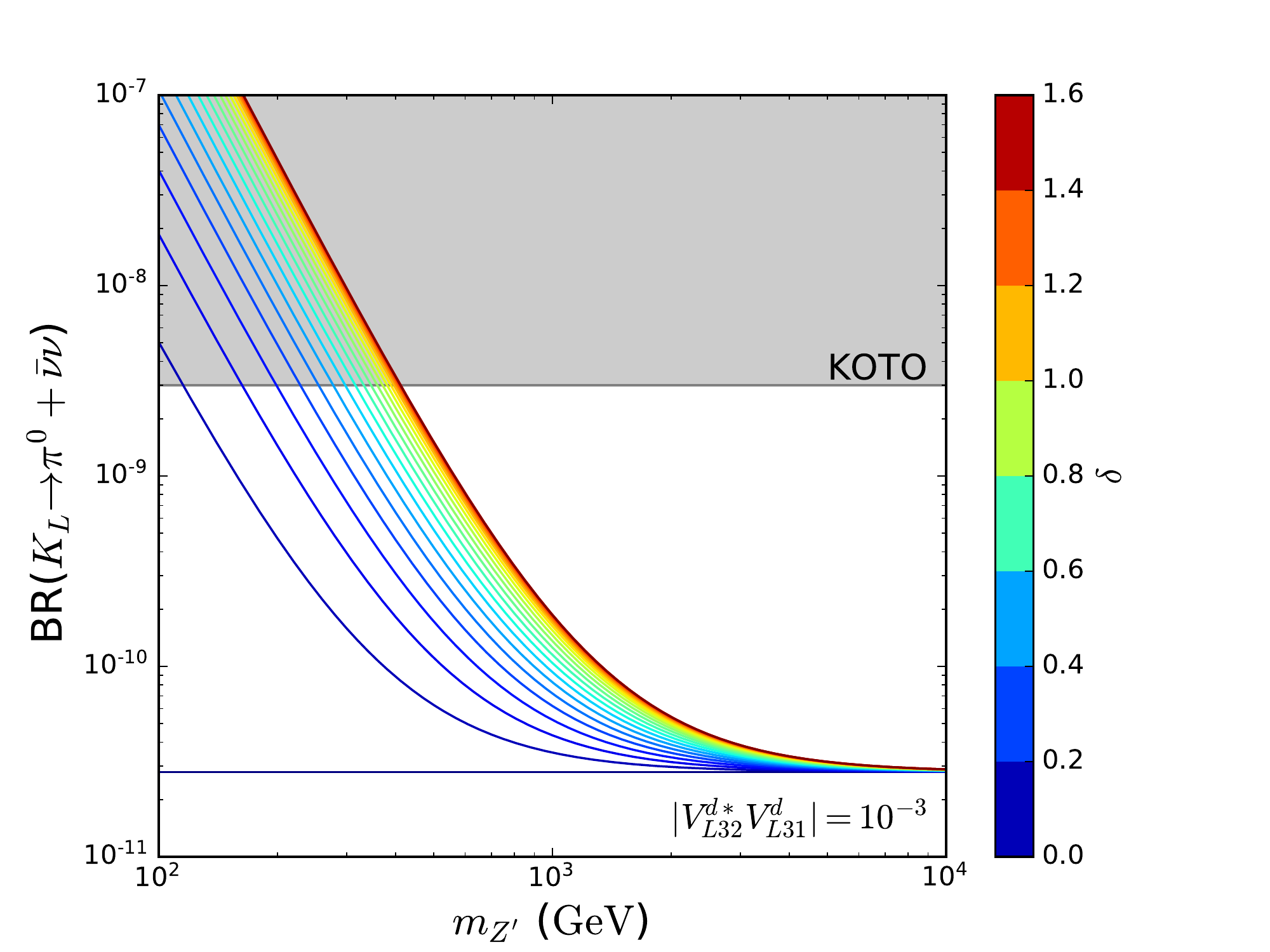}
    \includegraphics[width=0.48\textwidth]{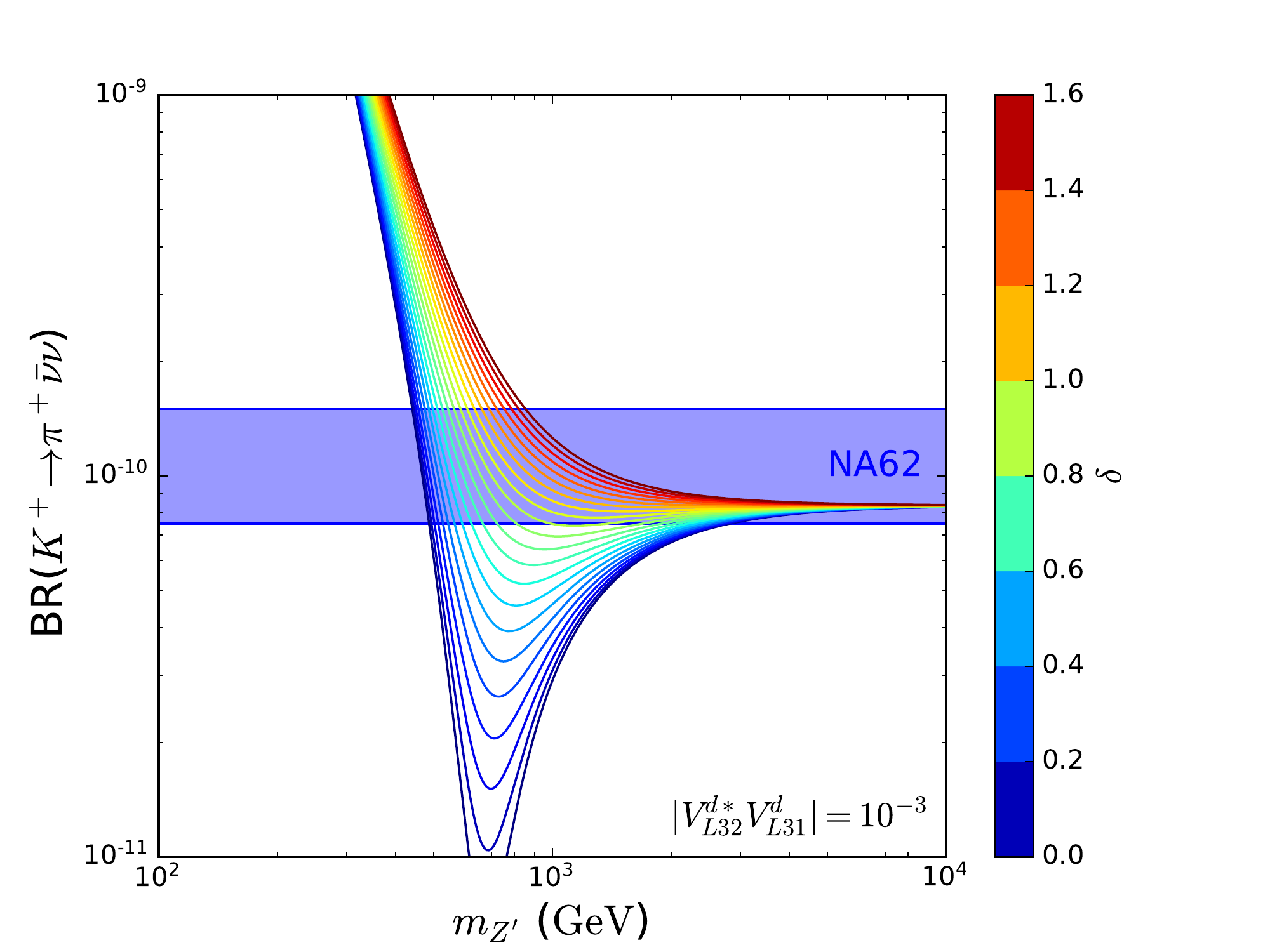}
    \caption{Branching ratio \textit{versus} $Z^\prime$ mass for kaon long (left panel), BR($K_L \to \pi^0 + \bar{\nu}\nu$), and for the $K^+$, BR($K^+ \to \pi^+ + \bar{\nu}\nu$) (right panel). The gray band provides the excluded region by KOTO experiment, and the blue region provides the allowed by the NA62 experiment. The colored bar represents the variation of the $\delta$ phase, for details see the text.} 
    \label{fig:BR}
\end{figure*}

\section{Results}
\label{sec:res}

In this section, we present our results for the FCNC processes using Eqs.~\eqref{eq:br_kp} and \eqref{eq:br_kL}, for the 331 r.h.n model, but we emphasize that our findings are also applicable to the 3-3-1LHN \cite{Mizukoshi:2010ky}, which is a 3-3-1 version that features heavy neutral leptons instead of right-handed neutrinos. Anyway, we compare our results with the last limits from KOTO \cite{Ahn:2018mvc} and, NA62 Run1 (2016 + 2017 + 2018)\footnote{\url{https://indico.cern.ch/event/868940/contributions/3815641/attachments/2080353/3496097/RadoslavMarchevski_ICHEP_2020.pdf}} \cite{NA62last,CortinaGil:2020vlo}. 
%
%
%
In Fig.~\ref{fig:BR}, we display the branching for $K_L$ (right panel) and $K^+$ (left panel) \textit{versus} the $Z^\prime$ mass for the 331 r.h.n model, namely combining the standard model contribution and the new one provided by the new $Z^\prime$ boson, as mentioned above. We compute both branching varying the $Z^\prime$ mass, the phase $\delta$ (color bar) and we fix the product $|V_{L32}^{d*}V_{L31}^d| = 10^{-3}$. From the left-panel of Fig.~\ref{fig:BR} one can conclude that the KOTO experiment excludes at most $Z^\prime$ masses around $400$~GeV, which occurs for $\delta > 1.4$. However, from the right-panel of Fig.~\ref{fig:BR}, we find that $Z^\prime$ masses below $3 \times 10^3$~GeV might be excluded depending on the value adopted for the phase $\delta$. For $\delta \rightarrow \pi/2$, the NA62 sensitivity to heavy $Z^\prime$ mediators severely weakens. Hence, NA62 yields complementary limits to other existing probes \cite{Lindner:2016bgg,Santos:2017jbv,Nepomuceno:2019eaz}. We can also see the presence of a dip due to destructive interference between the $Z ^\prime$ and the SM contributions, where the branching ratio lies bellow the SM predicted value. In the plot it occurs for $Z ^\prime$ masses in the $600-900$ GeV range, but in general its depth and location depends on the combination of $|V_{L32}^{d*}V_{L31}^d|$ and $\delta$. Notice that there is no dip when $\delta = \pi/2$, while it reaches its maximum depth for $\delta = 0$. \\

\begin{figure*}[htb]
    \centering
    \includegraphics[width=0.48\textwidth]{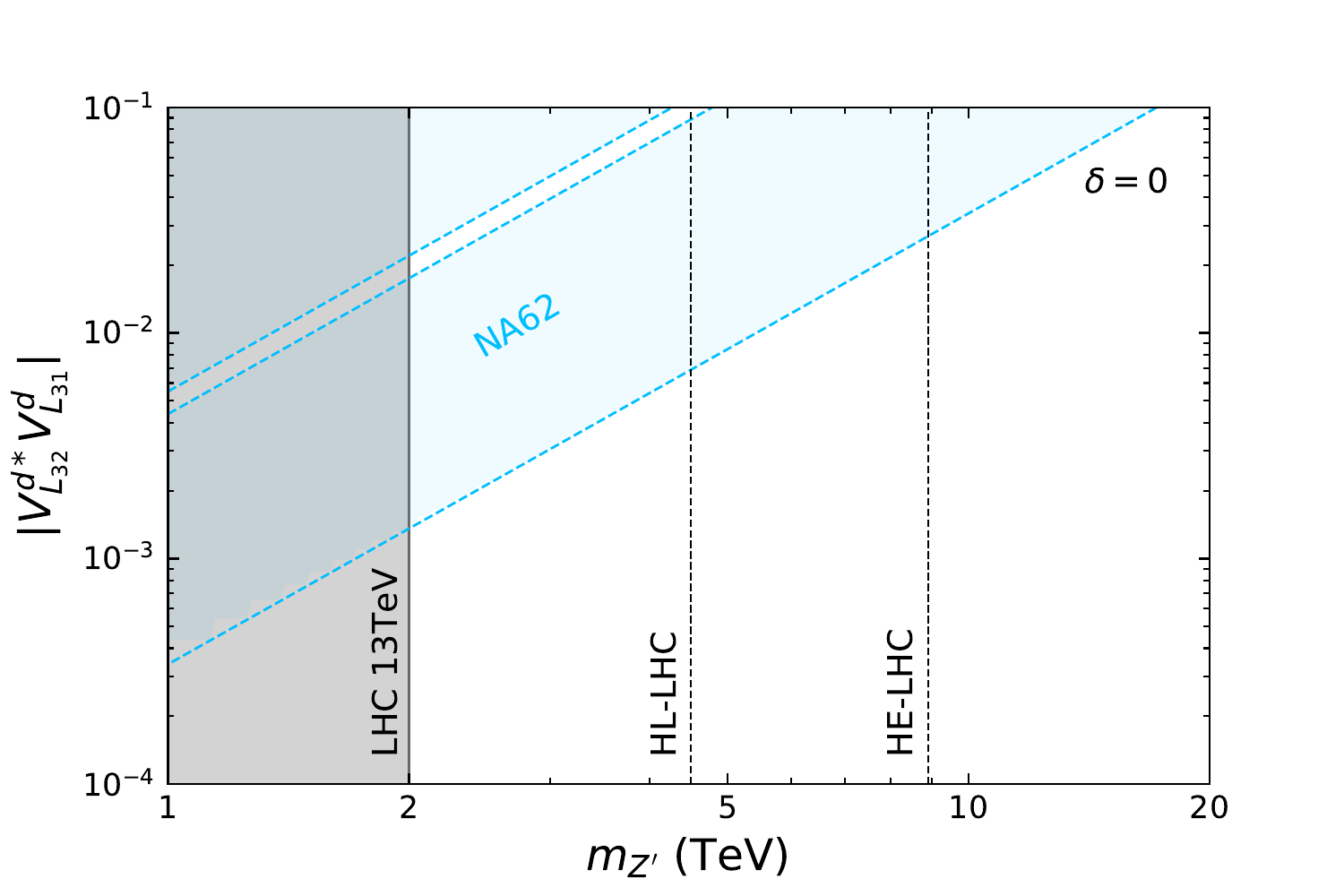}
    \includegraphics[width=0.48\textwidth]{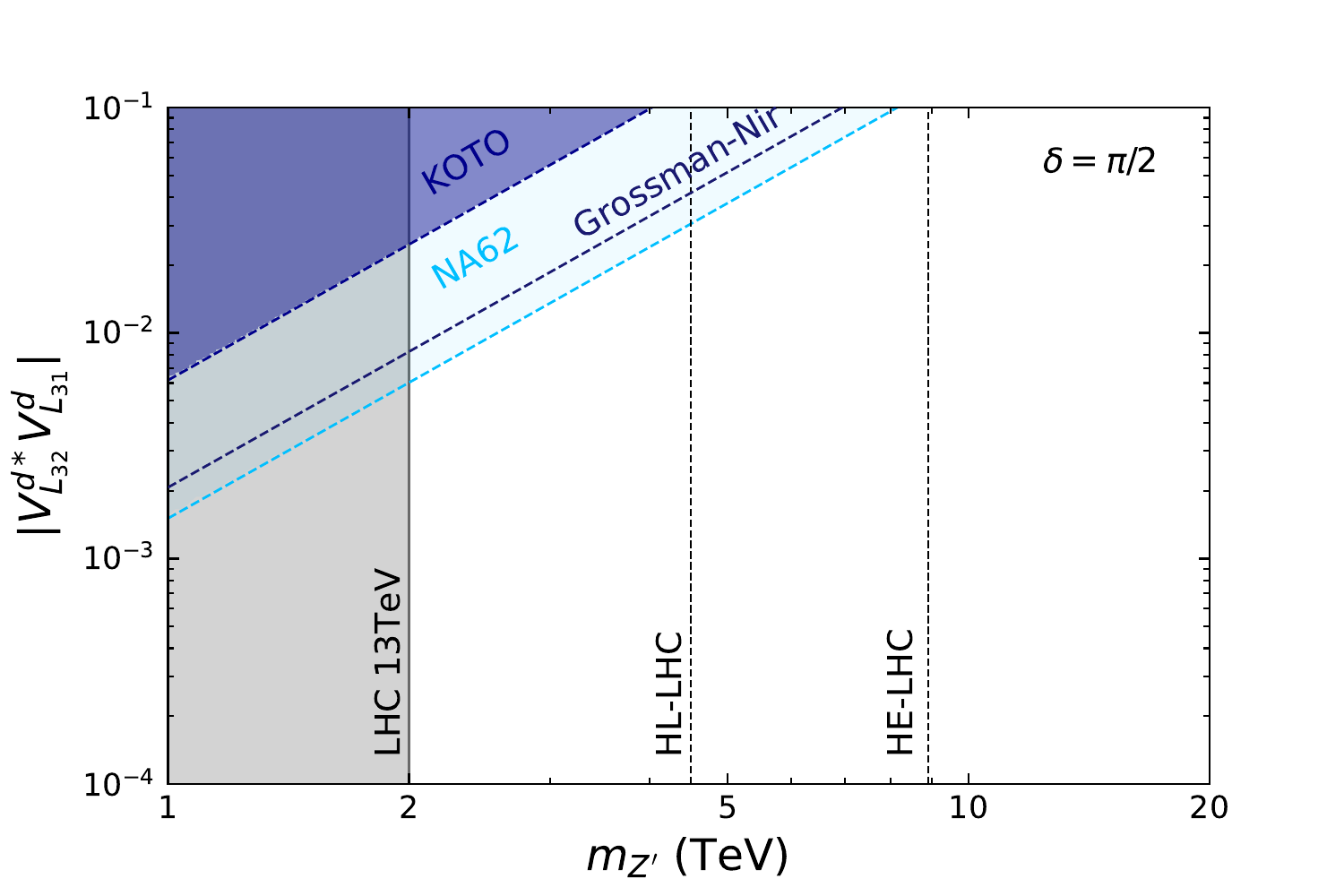}
    \caption{Excluded parameter space regions in the 331 r.h.n model, in the plane $|V_{L32}^{d*}V_{L31}^d|$ \textit{versus} $m _{Z^\prime}$, for $\delta = 0$ (left panel) and $\delta = \pi / 2$ (right panel). The colored regions represent the limits from NA62 (cyan), KOTO (blue) and LHC current (gray) and prospects (dashed lines).} 
    \label{fig:LHC}
\end{figure*}

We have drawn quantitative conclusions based on a particular value for $V_{L32}^{d*}V_{L31}^d$. To assess how our bounds change for different choices of this product we examine the range$10^{-4} < V_{L32}^{d*}V_{L31}^d < 10^{-1}$. 
In Fig.~\ref{fig:LHC} we show exclusion plots in the plane $|V_{L32}^{d*}V_{L31}^d|$ \textit{versus} $m _{Z^\prime}$, for fixed $\delta = 0$ (left-panel) and $\delta = \pi/2$ (right-panel). Here we 
combined limits from NA62, KOTO, and LHC, High-Luminosity LHC (HL-LHC) and High-Energy-LHC (HE-LHC). The strongest collider bounds on the $Z ^\prime$ mass stems from the resonant production of  $Z^\prime$ decaying into dileptons \cite{deJesus:2020ngn}.

There is an interesting interplay between collider and flavor physics. While collider bounds rely mostly on the interaction strength between the $Z^\prime$ boson and fermions, which is fixed in 3-3-1 models, the flavor bounds weaken with $|V_{L32}^{d*}V_{L31}^d|$. The collider bounds displayed in Fig.~\ref{fig:LHC} are conservative, as they take into account the presence of $Z^\prime$ decays into right-handed neutrinos and exotic quarks, which can be light enough so that the decays are not kinematically forbidden. The original lower mass bound reads $m _{Z ^\prime} > 4$ TeV \cite{Lindner:2016bgg}, but those exotic decays were ignored in \cite{Lindner:2016bgg}. If more decay channels become available, then the bound weakens. Here, in order to take into account this uncertainty on the $Z ^\prime$ decay modes, we assume conservatively a branching fraction of 50\% into charged leptons, which leads to the grey region in Fig. \ref{fig:LHC} in agreement with \cite{deJesus:2020upp}. We also show the prospects for the HL-LHC, with $3$ ab$^{-1}$ of integrated luminosity, and the HE-LHC, which corresponds to an integrated luminosity of $15$ ab$^{-1}$ at a center-of-mass energy of $27$ TeV. These projected collider limits were obtained using the code described in \cite{Thamm:2015zwa}, which can be used to forecast lower mass limits for resonance searches, which is precisely our case.

We found that the NA62 bounds from the decay $ K ^+ \rightarrow \pi ^+ \nu \bar{\nu} $ are rather restrictive, providing stronger limits than the ones from colliders over a significant region of the parameter space. Looking at left-panel of Fig.\ref{fig:LHC}, for instance, where $\delta = 0$, NA62 can exclude $Z ^\prime$ masses as high as $10$ TeV, if $|V_{L32}^{d*}V_{L31}^d|$ is of the order $10 ^{-2}$ or larger. In the same vein, NA62 enforces the product $|V_{L32}^{d*}V_{L31}^d|$ to remain below $\sim 10^{-3}$, when $m_{Z^\prime} \sim 2$~TeV.

On the other hand, these parameters are completely unconstrained by KOTO when $\delta = 0$, since the contribution from $Z ^\prime$ to the decay $K _L \rightarrow \pi ^0 \nu \bar{\nu}$ vanishes in this case. In the absence of new CP violating sources, 
the $BR (K _L \rightarrow \pi ^0 \nu \bar{\nu})$ takes the same value as in the SM, since this process will occur only through the contribution to CP violation from the SM via CKM matrix. 

This can be easily understood with Eq. (\ref{eq:br_kL}), from which we see that $BR (K _L \rightarrow \pi ^0 \nu \bar{\nu})$ depends only on the imaginary part of $X _{eff}$, in particular, on $Im (V _{ts} ^* V _{td})$ when $\delta = 0$.
Still in the left-panel of Fig. \ref{fig:LHC} we notice that in the range 
shown, the Grossman-Nir limit does not appear, because the suppressed values of $BR (K _L \rightarrow \pi ^0 \nu \bar{\nu})$ makes this bound easily satisfied for practically any reasonable value of $|V_{L32}^{d*}V_{L31}^d|$. However, as $\delta$ increases from $0$ to $\pi / 2$, this bound becomes more relevant, likewise the KOTO exclusion region gradually grows, reflecting the enhancement in the $K _L \rightarrow \pi ^0 \nu \bar{\nu} $ decay, while the NA62 exclusion region slightly decreases. 
Nevertheless, even with maximum enhancement at $\delta = \pi / 2$ (right plot), the KOTO bounds remain less constraining compared to NA62 and the Grossman-Nir limit. 

\section{Discussion}

Our conclusions relied on the presence of flavor changing interactions involving the $Z^\prime$ boson, but as the model features a large scalar sector, that could be new sources of flavor changing interactions rising from the heavy scalar fields. These new contributions have been shown to be subdominant. Thus can be safely ignored. Moreover, the entire reasoning was based on the 3-3-1 model with right-handed neutrinos, but our results are also applicable to the 3-3-1 model where the right-handed neutrinos are replaced by heavy neutral fields. This occurs because these models have the same neutral current. In summary, our finding are relevant for two different models and solid irrespective of the presence of scalar fields. 

\section{Conclusion}
\label{sec:conc}

In this work we explored the FCNC processes mediated by a $Z^\prime$ gauge boson featuring both in the 3-3-1 r.h.n and 3-3-1LHN models. We computed the $K ^+$ and $K _L$ decay rates to missing energy, considering the extra contributions from the $Z^\prime$ in addition to the SM contribution, leaving the quark mixing matrix and the $Z^\prime$ mass as free parameters. We performed a complementary analysis using the results from NA62, KOTO, and the LHC (current and prospects) to set bounds on the 331 r.h.n parameters. We found that the last result of the NA62 experiment was able to constrain a large region of the parameter space, setting lower limits on the $Z^\prime$ mass that can be more stringent than those from dilepton searches at the LHC. For example, we can impose $m_{Z^\prime} > 10$ TeV for $|V_{L32}^*V_{L31}| \sim 10^{-1}$, while $|V_{L32}^*V_{L31}| \lesssim 2 \times 10^{-3}$ for $m_{Z^\prime} = 3$~TeV. These results apply for $\delta=0$, where the sensitivity of NA62 is maximum. In the case when the new CP violation effects are large our constraints weaken. 

\acknowledgments

The authors thank Antonio Santos and Diego Cogollo for discussions. TM, CS and FSQ thanks UFRN and MEC for the financial support. FSQ also acknowledges the CNPq grants 303817/2018-6 and 421952/2018-0, the financial support from ICTP-SAIFR FAPESP grant 2016/01343-7, and the Serrapilheira Institute (grant
number Serra-1912-31613). FSQ and CS have been supported by the S\~{a}o Paulo Research Foundation (FAPESP) through Grant No 2015/15897-1. CS is supported by grant 2020/00320-9, S\~ao Paulo Research Foundation (FAPESP). SK acknowledges the support of the FONDECYT (Chile) grant No 1190845. Y. S. Villamizar acknowledges the ﬁnancial support from CAPES under grants 88882.375870/2019-01. This work was supported by the Serrapilheira Institute (grant number Serra-1912-31613). We thank the High Performance Computing Center (NPAD) at UFRN for providing computational resources.




\bibliography{main}

\end{document}